\documentclass[%
 reprint,
 amsmath,amssymb,
 aps,
]{revtex4-2}

\usepackage{graphicx}% Include figure files
\usepackage{dcolumn}% Align table columns on decimal point
\usepackage{bm}% bold math
\usepackage{subfig}
\begin{document}

\preprint{APS/123-QED}

\title{Single-shot wideband active mircorheology  using multiple-sinusoid modulated Optical Tweezers}% Force line breaks with \\

\author{Avijit Kundu\textsuperscript{1}}
\author{Raunak Dey\textsuperscript{1}}%
\author{Shuvojit Paul\textsuperscript{1,2}}
\author{Ayan Banerjee\textsuperscript{1}}
\affiliation{\textsuperscript{1}Department of Physical Sciences, Indian Institute of Science Education and Research Kolkata, Mohanpur Campus, Mohanpur, West Bengal, India}
\affiliation{\textsuperscript{2}Department of Physics, University of Konstanz, Germany}

\email{ayan@iiserkol.ac.in}

\date{\today}% It is always \today, today,

\begin{abstract}
We employ multiple sinusoid modulated optical tweezers to measure the frequency dependent rheological parameters of a linear viscoelastic fluid over five decades of frequency in a single shot, hitherto not achieved using active microrheology alone. Thus, we spatially modulate a trapped probe particle embedded in a fluid medium with a combination of a square wave - which is by definition a superposition of odd sinusoidal harmonics - and a linear superposition of multiple sinusoids at a wideband frequency range, with complete control over the amplitude, frequency and relative phase of the modulating signals. For the latter, we selectively excite the particle by larger amplitudes at high frequencies where the particle response is low, thereby enabling wideband active microrheology with large signal-to-noise. This mitigates the principal issue associated with conventional active microrheology - which is low bandwidth - and also renders our method better in terms of signal to noise, and faster compared to passive microrheology. We determine the complex viscoelastic parameters of the fluid by extracting the phase response (relative to input excitation) of the probe from the experimentally recorded time series data of the probe displacement, and employing well-known theoretical correlations thereafter. We test the efficacy of our method by studying a linear viscoelastic media (polyacrylamide-water solution) at different concentrations, and find good agreement of the measured fluid parameters with known literature values.
\end{abstract}

\maketitle

\section{\label{sec:level1}Introduction}
Complex fluids are ubiquitous in nature, and are essential to life itself due to their various diverse manifestations in biological entities, including the environments of cells themselves. In recent years, microrheology \cite{tassieri2019microrheology,squires2010fluid,furst2017microrheology} has proven to be an indispensable tool to study mechanical properties of complex fluids at microscopic lengths scales \cite{PhysRevE.81.026308,tassieri2015microrheology}. The technique generally employs microscopic probe particles in fluids, and the motion of these probe particles is measured to determine the fluid properties at different spatio-temporal scales. There are several experimental methods which are used for microrheological measurements including atomic force microscopy \cite{PhysRevLett.85.880,mackintosh1999microrheology}, optical tweezers \cite{tassieri2019microrheology, B907992K, Paul_2019}, and free Brownian diffusers \cite{PhysRevLett.74.1250}. The use of optical tweezers has specifically proven to be efficacious since they allow controlled manipulation of the probe particles inside a given fluid, which may be extremely useful in studying the heterogeneity in a biological sample \cite{WEIHS20064296}, tracking the local changes in a soft material over time \cite{B706004C}, as well as probing the interior of a cell \cite{lee2010passive}. 

Microrheology of complex fluids essentially consists of the determination of the complex shear modulus $G^{*}(\omega) [=G'(\omega)+iG''(\omega)]$, where the real part $G'(\omega)$ of the shear modulus  describes the elastic properties (storage), while the imaginary part $G''(\omega)$ describes the fluid damping properties (loss). Conventionally, microrheology is performed by tracking the Brownian motion of probe particles due to thermal (passive) motion using photo-detectors, diffuse wave spectroscopy, light scattering or video microscopy \cite{SAVIN2005623,CROCKER1996298,mason1997particle,Mason:97,dasgupta2005microrheology}, and using the statistics of the time series of the particles' displacement to determine the complex shear modulus $G^*{(\omega)}$ and viscosity $\eta^*{(\omega)}$ in the frequency domain, or stress $\sigma(\tau)$ and strain rate $\dot{\varepsilon}(\tau)$ for the time domain. Now, though the Brownian signal has the components of all the frequencies embedded in the time series, the signal power falls as $\frac{1}{f^2}$ ($f$ being the frequency), thus rendering passive mircorheology challenging at high frequency regimes. As a result, very often, long data-sets recorded over very long times are required to obtain acceptable signal-to-noise ratio at high frequencies. On the other hand, the response of a trapped probe particle can be enhanced by providing an external perturbation to it. Modulating the position of the minimum of the trap-potential sinusoidally (equivalent to an oscillating shear) \cite{C0CP01564D} converts the resultant motion of the particle into a superposition of Brownian and active motion. This approach leads to a much higher signal at the modulating frequency compared to the passive method, and is therefore known as active microrheology. \cite{doi:10.1021/ma801218z} The biggest issue with this technique is the long times required to obtain a data over a large bandwidth, since in that case the frequency of the perturbation needs to be varied continuously, with $G^*(\omega)$ measured for each frequency \cite{sriram2009small,ziemann1994local}. Any systematic changes that occur during that time in the trap (induced by laser drifts or fluctuations) or fluid properties, would introduce errors in the measured values. To address such issues, several techniques have been recently developed. These include flipping the probe particle between two time shared traps \cite{Preece2011}, using 'chirp waves' by continuously varying the modulation frequency \cite{PhysRevX.8.041042}, square waves \cite{Paul_2019} which contain odd sinusoidal harmonics, or even frequency modulation \cite{C5LC00351B} over a certain bandwidth. However, none of these single-shot active microrheology techniques are able to measure $G^*(\omega)$ for more than two to three 
decades of frequency \cite{choi2011active} (for examples, 0.01-10 rad/sec or 1-1000 rad/sec), so that low bandwidth remains a lacuna in single-shot active microrheology in comparison to the passive technique, where frequency responses have been collected for five decades in frequency, albeit with long measurement times and by employing a combination of techniques such as diffuse wave spectroscopy and light scattering from free probe particles \cite{mason1997particle,Mason:97,dasgupta2005microrheology}, or even active and passive microrheology performed in tandem using optical tweezers \cite{Preece2011}. Indeed, the broad frequency spectrum of thermally induced Brownian motion that is exploited in passive microrheology to garner viscoelasticity measurements over wide frequency ranges prompts the question whether this can also be achieved in active microrheology in some manner. This is exactly what we address in this work.

In this paper, we develop a new technique of active microrheology where we modulate a trapped colloidal probe particle using a ``Multiple Sinusoids Superposition Method''. Thus, we use a combination of square wave and a linear superposition of sinusoidal signals of known amplitudes, frequencies, and relative phases to modulate a single optically trapped probe particle to measure the complex viscoelastic parameters of a linear viscoelastic fluid in a single shot. We apply the square wave at low frequencies up to 500 rad/sec, and the superposition of sines at higher frequencies beyond 500 rad/sec. Earlier, we reported active microrheology using square wave excitation alone in Ref. \cite{Paul_2019}. Here, we develop the sinusoidal excitation technique, where - corresponding to our input excitation - the response of the particle is also a superposition of all frequencies, and we extract specifically the phase response at each excitation frequency by using a Discrete Fast Fourier Transform (DFFT) algorithm on the recorded time series of the particle's displacement \cite{7453}. Since the particle response falls off at high frequencies (compared to the corner frequency of the trap), we increase the modulation amplitude correspondingly, so that a substantial response is still available even at high frequencies. This innovation facilitates large signal-to-noise measurements even at high modulation frequencies \cite{Zhuang_2018}, thus enabling wide-band active microrheology. However, we take care to ensure that the modulation amplitudes are low enough to maintain the linear response of the viscoelastic fluid, so that the excitation at a particular frequency  elicits a response only at the same frequency, which we then carry out for all  frequencies of modulation.  We extract the complex viscoelastic parameters using a technique we reported first in our previous work \cite{Paul_2019}, and demonstrate our technique first on water for basic calibration, and then on a water-polyacrylamide (PAM) solution at different concentrations of PAM over a frequency range of $\sim 1.8 - 13900$ rad/sec, and observe that our measurements match literature values at similar domains of polymer concentration and frequency \cite{PhysRevE.71.021504,pommella2013using}. This technique thus mitigates a crucial issue of active microrheology - that of limited bandwidth - for linear viscoelastic fluids, and can also be evaluated for performance inside biological cells for measurement of rheological parameters. Also, to the best of our knowledge, high frequency rheology has not yet been performed in the case of biological entities. Thus, there is limited knowledge about the existence of interesting information at those regimes - which maybe a lacuna given that several cellular processes and  neural phenomena happen in the few milliseconds regime.

\section{\label{sec:level2}Microrheological analysis}

We model our system around the well-known \textit{Langevin} Equation for a spherical colloidal particle trapped in a harmonic potential created by a Gaussian beam in our case. We consider the fluid to be homogeneous around the locality of the particle throughout the experiment and assume that there is no active flow in our system. We modulate the minimum of the harmonic potential and detect the position of the particle in a single direction, so that the single variable differential equation in time for an independent degree of freedom can be written as:\\
\begin{equation}
m\frac{d^2x}{dt^2}=-\int_{-\infty}^{t}\gamma(t-t')\dot{x'}dt'-k[x(t)-x_0(t)]+\xi(t) \label{eq:1}
\end{equation}

Here $m$ is the mass of the spherical polystyrene particle, $k$ is the stiffness of the harmonic potential, $x_0(t)$ is the instantaneous position of the potential minimum, $x(t)$ is the position of the centre of the particle, and $\gamma(t)$ is  the time dependent friction coefficient of the viscoelastic fluid. The whole integral term on the right hand side of the equation calculates the drag force that acts on the particle, and has contribution from all past times. $\xi(t)$ represents the gaussian distributed white noise having an auto-correlation given by $\langle\xi(t)\xi(t')\rangle=2k_BT\gamma(t-t')$. Since the inertia of the particle is damped out by the fluid (the inertial time constant \cite{indei2012treating,cordoba2012elimination} of a few microseconds is much lower than the time resolution of our experiment), we ignore the inertial term of Eq.~\ref{eq:1} and write it in Fourier domain as,
\begin{equation}
[-i\omega\gamma(\omega)+k]x(\omega)=kx_0(\omega) \label{eq:2}
\end{equation}
Note that the useable range of frequency is up to around 100 krad/sec after which the inertia of the probe becomes significant \cite{waigh2005microrheology} (see Supplementary Information). The fluid being viscoelastic in nature has a complex drag coefficient, so that $\gamma(\omega)=\gamma'(\omega)+i\gamma''(\omega)$, and the phase response of the particle can be calculated as 
\begin{equation}
\tan(\phi)=-\frac{\omega\gamma'(\omega)}{k+\omega\gamma''(\omega)} \label{eq:3}
\end{equation}

The negative sign in Eq.\ref{eq:3} demonstrates that the phase response of the probe particle lags the excitation signal. Now, for two different values of $k$, say $k_1$ and $k_2$, we can extract two different values of the phase $\phi$, namely $\phi_1$ and $\phi_2$, respectively. For the potential minimum being modulated by $\omega_0$, we can then solve $\gamma'(\omega)$ and $\gamma''(\omega)$\\

\begin{equation}
\gamma'(\omega_0)=\frac{k_1-k_2}{\omega_0\left(\frac{1}{\tan(\phi_1)}-\frac{1}{\tan(\phi_2)}\right)} \label{eq:4}
\end{equation}

\begin{equation}
\gamma''(\omega_0)=\frac{\gamma'(\omega_0)}{\tan(\phi_1)}-\frac{k_1}{\omega_0} \label{eq:5}
\end{equation}

Since, $G^*(\omega)$, the complex shear moduli of the linear viscoelastic fluid is related to the drag coefficient of the fluid as $G^*(\omega)=-i\omega\gamma(\omega)/6\pi a$, where $a$ is the radius of the spherical probe, we can calculate the storage moduli (real part of G), loss moduli (imaginary part of G) and viscosity of the fluid as 

\begin{equation}
G'(\omega_0)=\omega_0\gamma''(\omega_0)/6\pi a \label{eq:6}
\end{equation}

\begin{equation}
G''(\omega_0)=\omega_0\gamma'(\omega_0)/6\pi a \label{eq:7}
\end{equation}

\begin{equation}
\eta(\omega_0)=\frac{\sqrt{G'(\omega_0)^2+G''(\omega_0)^2}}{\omega_0} \label{eq:8}
\end{equation}

Now, we modulate the potential minimum by a linear superposition of sine waves of known amplitudes ($A_i$), angular frequencies ($\omega_i$), and relative phases ($\psi_i$). So, 
\begin{equation}
x_0(t)=\sum_{i}A_{i}sin(\omega_i t+\psi_i)
\end{equation}
Using a Fourier decomposition on the final time series, each frequency component can be isolated and the respective phase and amplitude components extracted. The phase response of the particle (mentioned earlier as $\phi$) would be the difference between extracted phase $\phi^{ext}_i$ and input phase $\psi_i$, that is $\phi_i=\phi^{ext}_i-\psi_i$. For water, we can show that $\phi_i=\tan^{-1}\left(\frac{f}{f_c}\right)$ where $f_c$ is the corner frequency of the operating trap. $f_c$ is given by 
$$
f_c=\frac{k}{12\pi^2\eta_0 a}
$$
where $\eta_0$ is the viscosity of water. By fitting the phase response of the particle in water with varying frequency, we can find out the corner frequency and stiffness of our trap in water.\\

From Eq.~\ref{eq:2} we also calculate the amplitude response of the particle as
\begin{equation}
\mid x(\omega) \mid = \frac{kx_0(\omega)}{\sqrt{(k+\omega\gamma'')^2+\omega^2\gamma'^2}} \label{eq:10}
\end{equation}

For a simple, memory-less Newtonian fluid such as water,
$\gamma''=0$, and Eq.~\ref{eq:10} simplifies to \begin{equation}
\mid x(\omega) \mid = \frac{kx_0(\omega)}{\sqrt{k^2+\omega^2\gamma'^2}}
\end{equation}
Thus, the amplitude response of the particle is at a particular frequency, say $\omega_0$, is proportional to $\omega_0^{-1}$
\section{\label{sec:level3}Material and Methods}
\subsection{\label{sec:level1}Multisine method}
As mentioned earlier, we modulate the mininum of the harmonic trap with a linear superposition of multiple sine waves of different frequencies, amplitudes, and relative phase. We now describe our strategies about the choice of these parameters:\\
\subsubsection{\label{sec:level1}Choice of amplitudes}
\begin{figure}[h!]
	\centering
	\includegraphics[height=7cm]{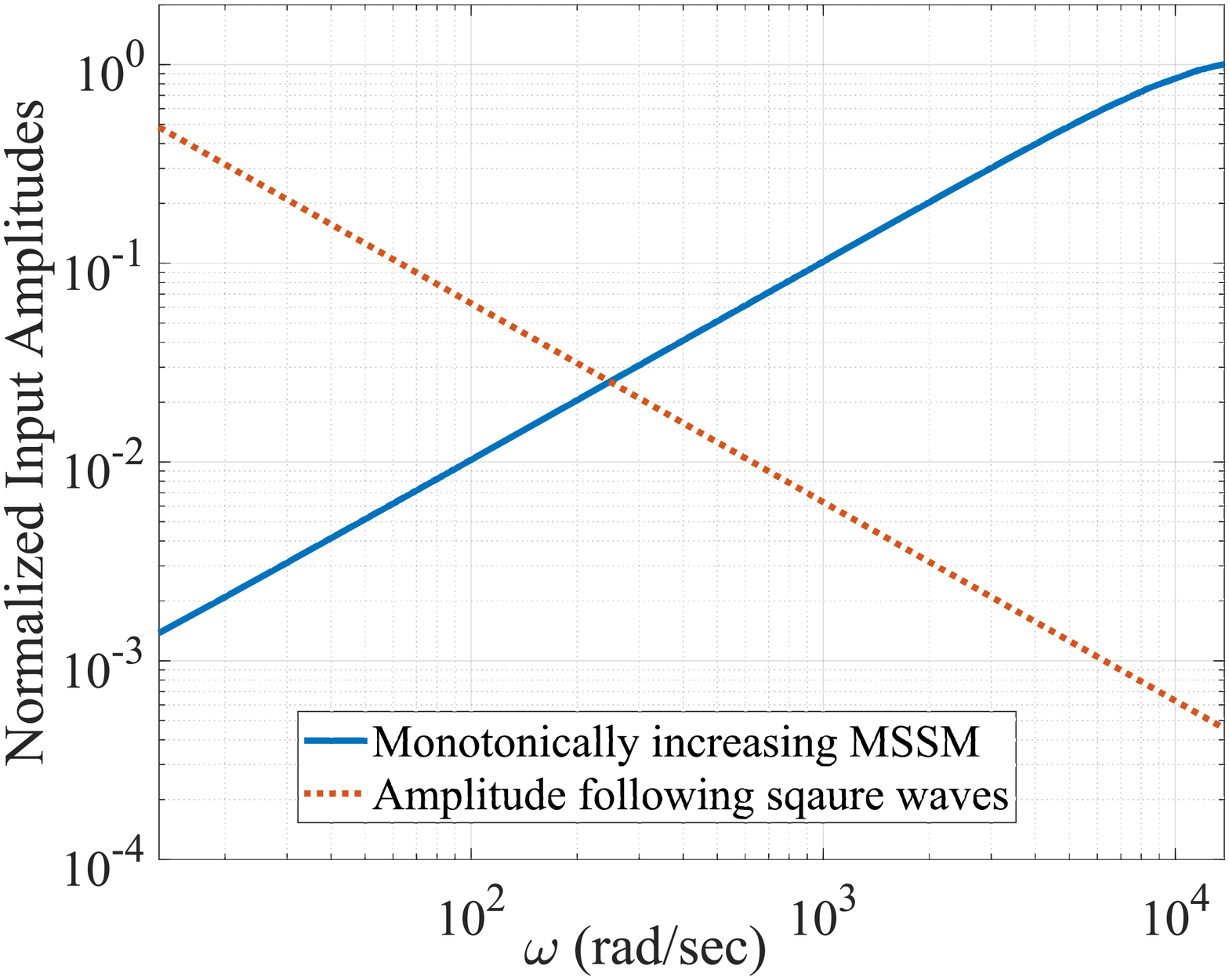}
	\caption{The normalised input amplitudes for two different kinds of multi-sines are plotted on a logarithmic scale against the input frequencies. We see that the monotonically increasing MSSM is better after $\sim250$ rad/sec than the multisine that follows square waves}
	\label{fig:amp_compare}
\end{figure}

We see from Eq.~\ref{eq:10} that the amplitude response of the probe particle is inversely proportional to the modulating frequency $\omega_0$. This poses a challenge in obtaining high signal-to-noise at high frequencies- therefore, we use input modulation amplitudes proportional to the input frequency. Our aim is thus to obtain a constant response over the entire frequency range of excitation in order to facilitate accurate phase and amplitude extraction. However, using only a monotonically linearly increasing input amplitude implies that the lower frequency components will be subjected to very low excitation, which will detrimental for phase extraction at these frequencies, especially for fluids with low concentrations. However, we have earlier shown \cite{Paul_2019}, that a square wave is essentially an infinite sum of the odd harmonics of sine waves mathematically written as
\begin{equation}
x_0(t)=\frac{4A}{\pi}\sum_{i=1}^{\infty}\frac{sin(2\pi(2i-1)ft)}{2i-1},
\end{equation}
where, $A$ is the amplitude of the square wave. As we see in Fig.~\ref{fig:amp_compare}, a square wave has large input amplitude at low frequencies with the amplitude decreasing linearly with increase in frequency. Thus, in our method of wide-band microrheology, we use a square wave excitation for estimating the complex viscoelastic parameters at the lower end of the frequency range,  and the increasing amplitude multisine excitation for the higher end of the frequency. Considering that we actually use the sinusoidal composition of the square waves for our method, we prefer to name this method of active microrheology as the \textit{'Multiple Sinusoids Superposition Method'} or \textbf{MSSM}. Our protocol finally consists of a single programmed sequence of pulses used to modulate the probe particle, which is made up of the following:
\begin{enumerate}
	\item Initially, we excite the particle for 100 seconds with a multi-sine signal in the form of a square wave, which forms a truncated series of odd sines with their amplitudes reducing with frequency. The frequencies we use are from 1-500 rad/sec roughly.
	\item  For the next 100 seconds we provide a second multi-sine signal where $A=constant\times \omega$, with $A$ and $\omega$ representing amplitude and frequency, respectively. The frequencies we use in this approach are from 500-14000 rad/sec.
\end{enumerate}
\subsubsection{\label{sec:level2}Choice of individual frequencies}
We can, in-principle, superpose any number of frequencies to obtain rheological information of the fluid at great detail in frequency space. But the superposed signal needs to be normalised to a certain peak-to-peak amplitude such that when it is applied to the particle, the oscillation of the particle remains constrained in the harmonic region of the trap, where the linear response approximation is valid \cite{PhysRevE.97.042606}. To choose the values of the frequencies, we choose a series of prime numbers in order to ensure that the response at one frequency is not mixed up in the next harmonic (though we are working in the linear region of the VE fluid, where this should not occur in theory). Also we observed that using prime-numbers minimizes the chances of constructive interference in our input waveform, thus create stronger signal at each individual frequency after normalisation. The frequency values $f_p^i$ are therefore chosen as  $f_p^i<f_{max}$, where $p$ is a prime number and $i$ is a positive integer. In our case $f_{max}$ is 2197 Hz (13.8 Krad/sec). This frequency is close to an order of magnitude higher than the corner frequency of our traps at two different powers and it is important to note that the effects of the trap diminish rapidly at even higher frequencies. Hence, we treat this frequency as a cut-off (note that the noise intensity goes as $power\sim\frac{1}{f^2+f_c^2}$, and thus diminishes by two orders of magnitude of the mean intensity if the operating frequency is increased by one order).
We also observed that the input phase has no significant effect on the accuracy of the phase extraction. We have therefore attempted experiments using linearly increasing phase difference, zero phase difference, as well as random phase difference between individual components, but have not observed any modifications in the nature of the results we obtained. 

Finally, we conclude that by combining two different genres of \emph{multisines} for equal time intervals, we are able to create a custom waveform of duration over 200 seconds and perform single-shot wide-band microrheology.

\section{\label{sec:level4}Experimental design}
\begin{figure}[h!]
	\centering
	\includegraphics[width=8.5cm]{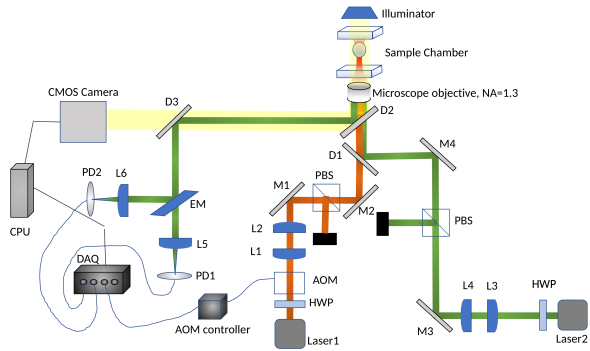}
	\caption{The schematic of our setup is presented here. The annotations are as follows: HWP - half wave plate, AOM - acousto optical modulator, L - convex lens of various focal lengths, PBS - Polarizing Beam Splitter, D - Dichroic mirror, M - one sided mirror, PD - photodiode, DAQ - data acquisition card (manufactured by National Instruments.)}
	\label{fig:setup}
\end{figure}

We develop our optical tweezers setup (Fig.~\ref{fig:setup}) around an inverted microscope (Olympus IX71) with an oil immersion objective lens (Olympus 100x, 1.3 numerical aperture) and a semiconductor laser (Lasever, 500 mW max power) of wavelength 1064 nm which is tightly focused on the sample. We modulate the beam by using the first order diffracted beam off an acousto-optic modulator (AOM), (Brimrose) placed at a plane conjugate to the focal plane of the objective lens. The modulation amplitude is small enough such the power in the first order diffracted beam is modified very minimally (around 1\%) as the beam is scanned.  We employ a second stationary and co-propagating laser beam of wavelength 780 nm and very low power (less than 5\% of the trapping laser power, such that the detection laser does not influence the motion of trapped particle in any way) to track the probe particle's position (detection laser), which we determine from the back-scattered light that is incident on a balanced detection system \cite{doi:10.1063/1.3685616}. The balanced-detection-system together with a data acquisition card record the probe displacement data into a computer. We prepare a sample chamber of dimension around $20$ mm$ \times$ $10$ mm$ \times$ 0.16 mm by attaching a cover slip to a glass slide by a double-sided tape which contains our model viscoelastic fluid. The model fluid samples are prepared by mixing polyacrylamide (PAM, flexible polyelectrolytes, $M_w = (5 - 6) \times 10^6$ gm/mol, Sigma-Aldrich) into water along with mono-dispersed spherical polystyrene probe particles of radii $1.5$ $\mu$m in very low volume fraction ($\approx0.1\%$) which we use as probes. We also prepare the sample just before performing our experiments so that there are no ageing issues associated with PAM. We use PAM solution at 5 different concentrations in water: 0.01\% w/w, 0.05\% w/w, 0.1\% w/w, 0.5\% w/w and 1\% w/w.

For each viscoelastic sample, first, we trap a single probe particle around 30 $\mu$m away from the nearest wall to get rid of any surface effects \cite{kundu2019measurement} and record its Brownian motion with sampling frequency 5 kHz over 20 s, to measure the stiffness of the trap. Then, we modulate the trap center by a square signal of very small peak-to-peak amplitude (130 nm) - so that the particle always remains in the linear region of the trap. The fundamental frequency of the square waves are set to 0.3 Hz (or 1.8 rad/sec) for all the cases. To perform the experiments using the multisine excitation, we follow the same protocol, only this time we use 100 seconds of observation for the multi-sine signal keeping the sampling frequency fixed at 5 kHz. To achieve a high degree of precision we maintained the experimental setup unaltered during the experiment. The rheological parameters can depend on the trap stiffness, probe size, and age of the sample-- for which we have used freshly prepared solutions of PAM everytime we performed an experiment.

\section{\label{sec:level5}Results and Discussions}
Equilibrium statistics allows us to measure the stiffness of the optical trap using Maxwell's Equipartition Theorem. This is free from the rheological influence of the sample used \cite{PhysRevE.81.026308,neuman2004optical}. According to this theorem, the trap stiffness is given by, $k=k_BT/\langle (x-\langle x\rangle)^2\rangle$. Since the histogram of the positions follows a gaussian distribution, we find out $\langle (x-\langle x\rangle)^2\rangle$ by fitting the data. That gives us an accurate measure of $k$, for all the viscoelastic samples we have used in our experiment. Before using the equipartition method to find out the trap stiffnesses at two different powers, we need to find out the sensitivity of our position detection system. Note that position calibration is independent of the sample rheology. We have used the viscous drag method in water to find out the corner frequency at two different powers and calibrations of our setup. We use this method to calibrate our optical trap from the voltage (as obtained in our detector) to particle displacement. We use the fact that the phase lag of the particle response in water can be accurately fitted with $\tan^{-1}\left(\frac{f}{f_c}\right)$, which gives us the value of $f_c$ - the corner frequency. For two different powers of the trapping laser we calculate the corner frequency as $116.6\pm7.4$ Hz and $231.1\pm20.0$ Hz, for low and high trapping powers, respectively. From this, we find out the trap stiffness from $k=12\pi^2\eta_0 af_c$, which leads to trap stiffnesses of $17.61\pm1.12\ \mu N/m$ and $34.87\pm3.02\ \mu N/m$, respectively in water.

However, it should be noted that our method is built on a phase extraction algorithm - we do not use the amplitude component due to the associated signal-to-noise issues during extraction, especially for low concentrations of the viscoelastic (PAM) component \cite{Paul_2019}. Nonetheless, we perform the spatial calibration to determine the trap stiffness and to ensure that our trap modulation leads to particle displacements that lie within the linear response of the trap. We also check that the corner frequency changes linearly with power so that we know that the trapping potentials we use are indeed harmonic.

It is important to note that for extraction of the phase response from the data, trap calibration is not required. Indeed, our procedure for phase extraction is entirely based on the \emph{Discrete Fast Fourier Transform method}. We take the Fourier transform of the entire time series data of length $N$ and in accordance with the Nyquist criterion, dispose off half the data - so that we are left with a complex time series of $N/2$ points. The ratio of the imaginary to real components of each number in the Fourier transformed time series gives the tangent of the phase value. Our task is essentially to find the phase corresponding to the excitation frequency we are interested in. In a similar vein, the amplitude response can be calculated as the modulus of each number. The time for the analysis to run is highly dependent on the time taken to compute to Fourier Transform, because rest of the operations such as calculating rheological parameters from extracted phases have linear time complexity. As a result, the time complexity of our entire method follows that of a \emph{Fast Fourier Transform} as $\sim O(NlogN)$. Also, the high frequency limit of our experiment is set by the corner frequency of our trap - which presently - at the highest laser power we can couple into the trap, is around 232 Hz. Increasing both trap stiffness and sampling frequency will thus facilitate the increase of the high frequency limit of this technique to whatever level is desirable.

All the individual components of the multi-sine wave affect the probe particle independently. We initially verified that we could extract individual components using the DFT algorithm for a multisine wave, and also matched our results with experiments using single sine waves of different frequencies. Most importantly, we ensured that there was no crosstalk - i.e., no leaking of the response at one frequency into another, which implied that the displacement due to a particular force at a particular frequency was linearly related to the response of the system. Thus, for the displacement $x_o(t)$ by the input perturbation $x_i(t)$, we have  $x_o(t)=\int dt' \chi(t;t') x_i(t')$. Our system - comprising of the trapped particle in a viscoelastic fluid gives - a linear response $\chi$ under small perturbations of the particle. This response function is invariant under time translation, $\chi(t;t')=\chi(t-t')$, so that performing a Fourier transform of $x_o(t)$, we obtain

\begin{equation} \label{eq13}
x_o(\omega)=\int dt' \int dt e^{i\omega t}\chi_{oi}(t-t')x_i(t') = \chi_{oi}(\omega)x_i(\omega)
\end{equation}

This essentially is the general form of a linear response which we also see in Eq.~\ref{eq:10} for our special case, and this implies that the perturbation is `local' in the frequency domain. In other words, the output response of the particle is linearly independent at different frequencies for the small displacements  (of the order of tens of nanometers) we drive in our technique. In order to verify that quantitatively, we determine the velocity of the particle $x_i(t)$ using Euler's differential method. We use this extracted velocity to compute the Weissenberg number, and observe that in all our experiments, we have $Wi << 1$. This indicates that we are indeed eliciting linear response from the particle as a result of the effects on it from the viscoelastic fluid \cite{gomez2016dynamics,PhysRevLett.113.098303,paul2019active} (see Online Supplementary Information). We would like to emphasise that this estimation is crucial - especially for soft materials having low yield strain - in which case a trade-off between the linearity and magnitude of response required to perform a reliable measurement may lead to a change in the frequency range over which the viscoelastic properties can be measured - especially at the higher end.

In order to experimentally validate our algorithms for phase and amplitude extraction, we perform experiments in water, and observe that while $G'(\omega)$ is almost zero for the entire range of modulation frequencies, $G''(\omega)$ follows the trend expected for water theoretically. This is shown in Fig.~\ref{fig:moduli}b. We now move on to the PAM-water mixtures, and determine the amplitude and phase responses of the trapped probe, which we then compare to determine the parameter that would be the most efficacious to use to measure the complex viscoelasticity of our samples. Now, the amplitude response of the particle is defined as $\frac{x_o(\omega)}{kx_i(\omega)}$, where, $x_i(\omega)$ and $x_o(\omega)$ are the Fourier components of input signal provided to the probe particle, and output signal from the particle as measured using photo-diodes (both transformed in distance units, respectively). This amplitude response can be theoretically calculated as shown in Eq.~\ref{eq:10}, but we do not employ it to extract the rheological parameters of the fluid as mentioned earlier. However we note that, for higher concentration samples the amplitude response is also higher (See Online Supplementary information), as a result of which,  we are able to extract the whole broadband range of rheological parameters for the 0.1\% w/w, 0.5\% w/w and 1\% w/w PAM in water samples with our MSSM technique {\it only (without the aid of square wave modulation)}. For lower concentration samples, signal-to-noise issues prevent us from obtaining reliable values for $G'(\omega)$ and $G''(\omega)$.

\begin{figure}[h!]
	\centering
	\includegraphics[height=6cm]{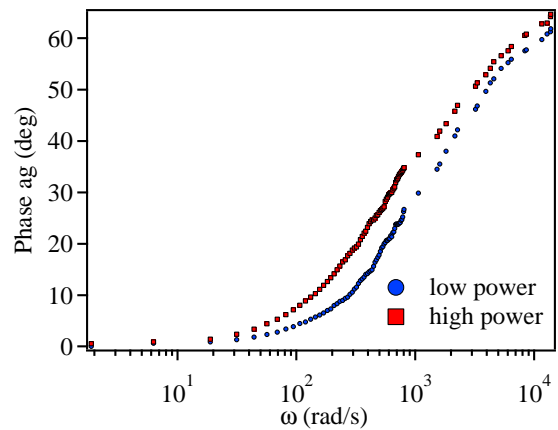}
	\caption{Phase response of the trapped probe particle embedded in the 0.01\% w/w sample to the Multisine perturbations plotted over the entire frequency range in a semi-logarithmic graph. The two different curves are obtained for different sets of measurements using two different powers of the trapping laser.}
	\label{fig:phase}
\end{figure}

\begin{figure}[h!]
	\centering
	\subfloat[Storage moduli]{{\includegraphics[height=6cm]{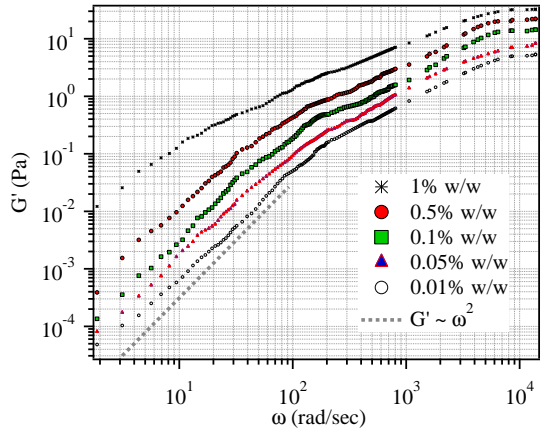} }}%
	\qquad
	\subfloat[Loss moduli]{{\includegraphics[height=6cm]{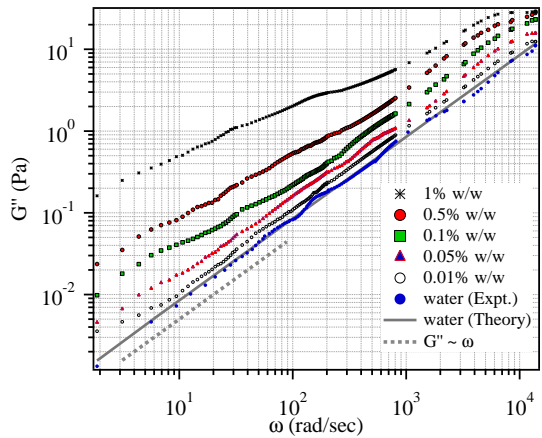} }}%
	\caption{The storage moduli (a) and loss moduli (b) for 5 different concentrations of PAM in water, (1\% w/w, 0.5\% w/w, 0.1\% w/w, 0.05\% w/w and 0.01\% w/w) extracted using our technique and plotted against angular frequency. We also have plotted the loss moduli of water  using our method along with theoretical values. As expected we see a gradual increase in both of them with increasing frequency. Trendlines to confirm that $G'$ and $G"$ scale as $\omega^2$ and $\omega$, respectively, at low frequencies where $G">G'$, are also plotted in grey in each figure.}%
	\label{fig:moduli}%
\end{figure}

\begin{figure}[h!]
	\centering
	\includegraphics[height=6cm]{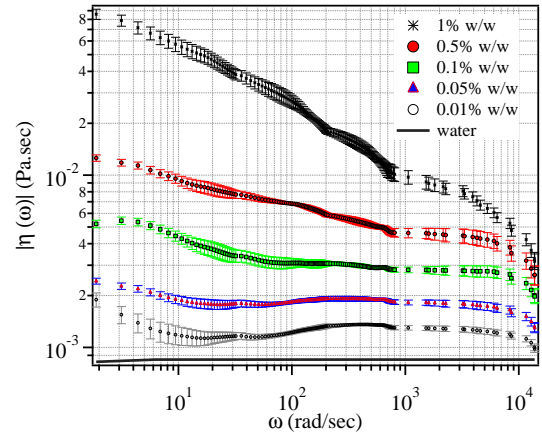}
	\caption{Viscosity for 5 different concentrations of PAM in water, (1\% w/w, 0.5\% w/w, 0.1\% w/w, 0.05\% w/w and 0.01\% w/w) extracted using our technique and plotted against angular frequency; standard error, obtained after averaging over 5 data sets, is shown as the error bar. We observe a gradual decrease in the effective viscosity of the sample with increasing frequency of oscillations, and after a cut-off frequency it reaches an asymptotic value.}
	\label{fig:eta}
\end{figure}

\begin{figure}[h!]
	\centering
	\includegraphics[height=11cm]{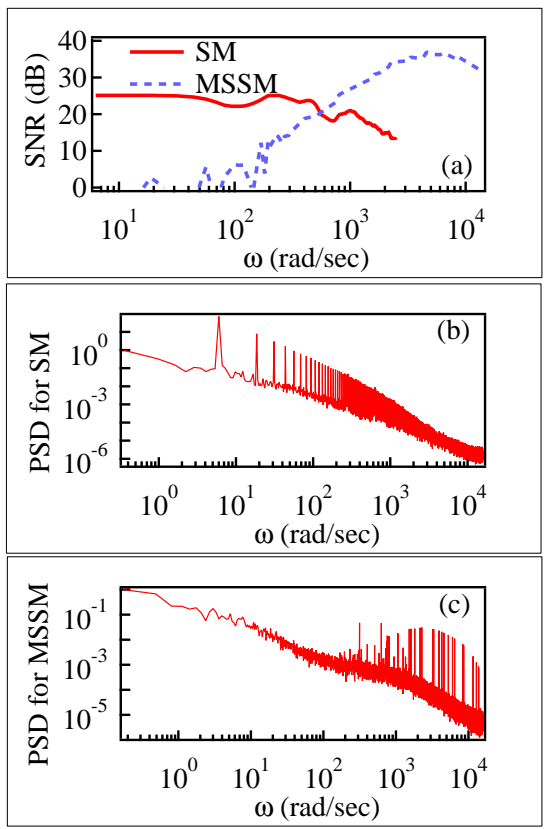}
	\caption{(a) Signal-to-noise ratio over the entire frequency range, where (b) is the power spectrum density (psd) from square wave modulation (SM), which shows the amplitude response to be significantly high  up to 500 rad/sec and (c) is the psd from MSSM depicting the amplitude response to be significantly high after 500 rad/sec. All the graphs are plotted for the dataset 0.1\% w/w PAM in water, just taken as an example.}
	\label{fig:snr2}
\end{figure}

\begin{figure}[h!]
	\centering
	\includegraphics[height=5.5cm]{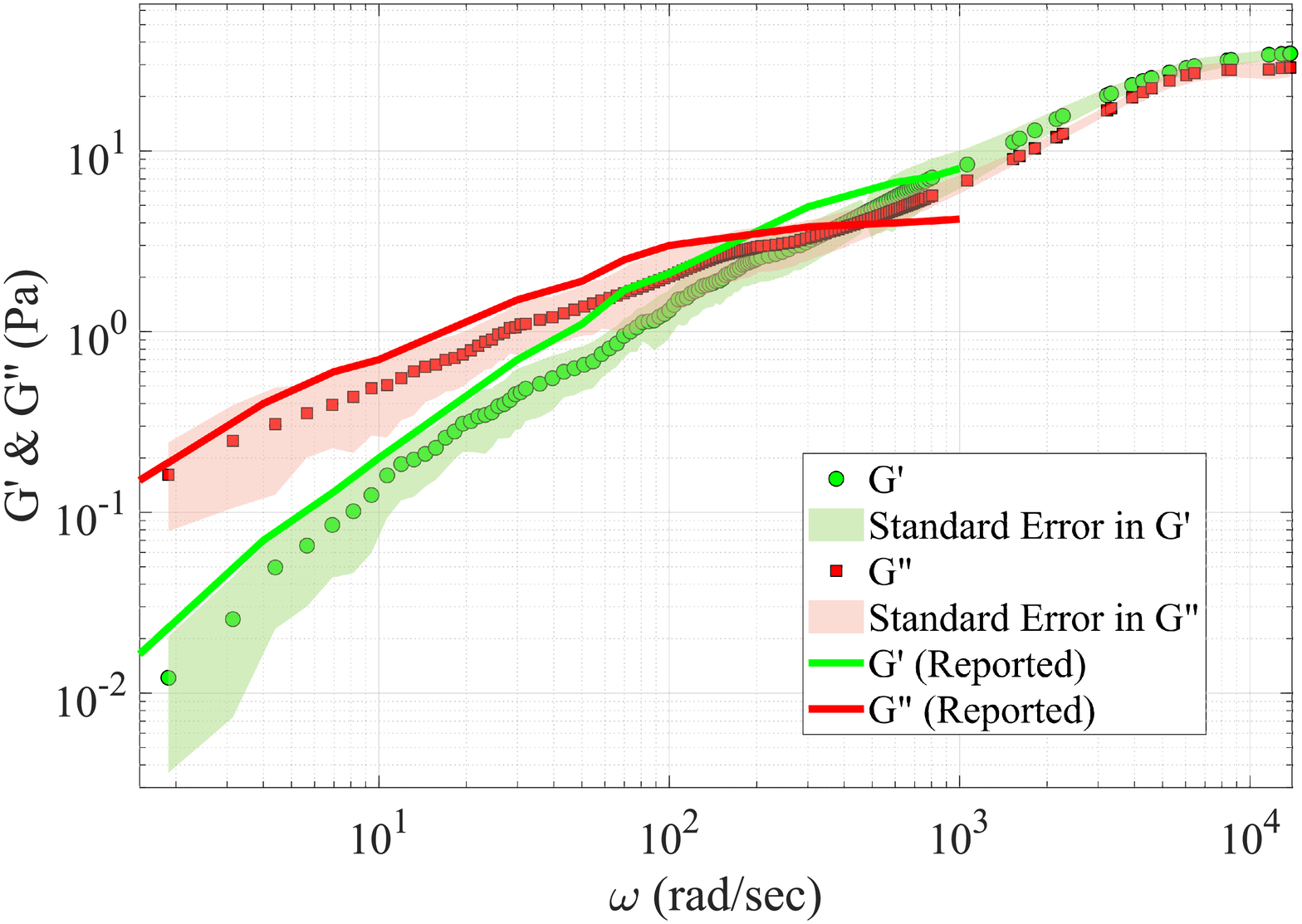}
	\caption{The storage and loss moduli ($G'$ and $G"$) are plotted in green squares and red dots respectively for 1\%w/w PAM in water using the protocol described in this paper. To compare we show the values at the same concentration for a smaller particle size as reported in Fig.5 of Tassieri \textit{et al.} (Ref.~\cite{PhysRevE.81.026308}). We observe that the measured values closely concur with the reported values, which further corroborates our technique.}
	\label{fig:moduli_compare}
\end{figure}

We now analyse the phase response of the probe more carefully. In Fig.~\ref{fig:phase}, we observe that the phase values for higher trap stiffness is less than the phase values for the lower stiffness. This is understandable, since the phase response is inversely proportional to corner frequency $\left(\tan^{-1}\left(\frac{f}{f_c}\right)\right)$. We extract the phase, as shown for one sample in Fig.~\ref{fig:phase}, for 2 different powers for each concentration. Using the measured phases and using Eqs.~4-8, we calculate the storage and loss moduli for each sample. These are plotted in Fig.~\ref{fig:moduli}. It is clear that with the increase in frequency of the probe particle - which means an increase in its velocity through the fluid - both the storage and loss moduli of PAM increase. This is because -- as the probe oscillates at higher frequencies -- it deforms the polymeric mesh progressively, and also significantly enough for the latter to affect the motion of the particle, and thus alter its response to the drive. For low drive frequencies, the mesh is able to relax - so that the particle does not see its effect. Similarly it is easy to understand that with the increase in concentration of PAM in water, the overall density of the mesh would increase, so that it would be able to impart more elastic energy to the probe, as we observe in Fig.~\ref{fig:moduli}(a). In the same manner, the loss also increases since dissipation is higher now, and we also observe in Fig.~\ref{fig:moduli}(b), that all the measured values of $G''$ lie above the theoretical value of $G''$ in water as expected (for water, a purely viscous fluid, $G''(\omega)=\eta\omega$ and $G'(\omega)\sim 0$, since there is no polymeric constituent for storage).  The effective viscosity is therefore higher than that of water, and only approaches that value asymptotically at high frequencies, as is clear from Fig.~\ref{fig:eta}. For each of the concentrations we have performed multiple measurements, and averaged over 5 data sets to ensure that our results are reliable. In Fig.~\ref{fig:moduli}(a) and (b) only the mean is plotted in the graph, though, to avoid unnecessary congestion. However, in Fig.~\ref{fig:eta}, both the mean and the standard error in the measured values of the viscosity have been plotted. In addition, as expected from the theory, the shear moduli follows the Kramers-Kroenig relations, which is demonstrated by the fact that $G'$ and $G"$ scale as $\omega^2$ and $\omega$, respectively, at low frequencies where $G">G'$. We explicitly show this in Fig.~\ref{fig:moduli}.\cite{mason1997particle}.

An interesting observation in Fig.~\ref{fig:moduli}(a) is that $G'$ saturates beyond a particular frequency. Such saturation as a function of frequency has been observed for entangled polymers in several other cases \cite{tassieri2019microrheology,PhysRevE.81.026308,Preece2011,LIU201746,Paul_2018,chandra2018onset}. We believe that this marks a transition from the polymer from the terminal region to the rubbery plateau region as is described succinctly in Ref.~\cite{LIU201746}. Alternatively, it has also been understood as the saturation of $G'$ beyond the frequency corresponding to the polymer time constant \cite{Paul_2018,chandra2018onset}. The time constant increases for increasing concentrations - as a result of which the saturation frequency also lowers correspondingly. We also observed that the saturation effect for the highest concentration (1\% W/W) was not visible at the modulation amplitude we used for other concentrations, i.e. around 130 nm. Reducing the amplitude to 70 nm, however, led to saturation of $G'$ as is visible in red solid line with star-shaped data points in Fig.~\ref{fig:moduli}(a). This is because the condition of $W_i<<1$ required for the linear viscoelastic regime is not enforced strongly enough for the modulation amplitude of 130 nm (see more details and a plot of $G'$ for two modulation amplitudes in the Supplementary Information). 

It is also interesting to quantitatively compare the signal-to-noise ratio for a square wave and the multi-sine excitations, as we demonstrate in Fig.~\ref{fig:snr2}(a) - (c). To generate the data shown in this figure, we actually excite the probe by a square wave excitation, and a multi-sine wave till 13982 rad/sec. The signal-to-noise is defined as the amplitude of detected response above the thermal noise level, as obtained from the power spectral density. We define it in decibel units as: $$S/N=10\log_{10}\left(\frac{\text{Power of signal}}{\text{Power of noise}}\right)$$. Thus, we consider the noise power spectrum of the Brownian motion and determine the amplitude of the response peak we obtain at a particular excitation frequency, and its strength with respect to the noise floor defined by the pure thermal motion. We determine the signal-to-noise values for both square and multi-sine wave excitation and compare them in Fig.~\ref{fig:snr2}(a). Fig.~\ref{fig:snr2}(b) and Fig.~\ref{fig:snr2}(c) shows the peaks in the power spectrum of square and multisine respectively for the case 0.1\% w/w PAM in water as an example.  We observe that at angular frequencies less than $\sim500$ rad/sec, the square wave gives a higher signal-to-noise ratio for the particle response - however beyond that, the multi-sine wave gives continuously increasing signal-to-noise ratio. This motivates us to use a square wave excitation till 500 rad/sec, and a multi-sine excitation beyond that frequency. Another issue we consider is the continuity in phase retrieval between the square wave excitation and multi-sine excitation. Even here we observe that there is indeed a smooth continuity between the extracted phase by square wave and that by the multi-sine approach. This is borne out from the fact that the extracted complex rheological parameters  change smoothly, as is clear from Fig.~\ref{fig:moduli}, where the data for frequencies less than 500 rad/sec is from a square wave excitation, while that for higher frequencies is via the multi-sine excitation. Finally in Fig~\ref{fig:moduli_compare}, we compare the viscoleastic parameters that we have extracted with that measured in the literature in Fig.5 of Tassieri \textit{et al.} (Ref.~\cite{PhysRevE.81.026308}). We plot the mean values of G' and G" from multiple measurements and provide the standard error obtained after averaging over 5 data sets. It is important to note that the size of the probe particle used in Ref.~\cite{PhysRevE.81.026308} is $5 \mu m$, which is greater in size than the $3 \mu m$ particles we use. Although the trend matches with the reported literature value, the systematic deviation is probably due to the difference in particle size, which may suggest that PAM has multiple different length scales and is a multi-scale mesh \cite{weigand2017active}. 

The first crossover between the two shear moduli gives the time constant of the fluid, from the inverse of the crossover frequency (which is $\approx 5$ ms in our case for 1\% w/w PAM in water). Presence of multiple crossovers of the shear moduli may suggest that PAM has multiple polymeric time constants emerging from different length scales (multi-scale mesh). In Fig.~\ref{fig:moduli_compare}, we observe that the values of $G'(\omega)$ and $G''(\omega)$ seem to approach each other - but it is difficult to infer conclusively about the possibility of a crossover happening at even higher frequencies, since the inertial effects of the medium start showing up \cite{tassieri2019microrheology} at that regime, which may complicate the physics considerably. For a discussion on the inertial effects in microrheology see the Online Supplementary Information. Clearly, more experiments with different probe sizes are called for to clarify this matter.

\section{\label{sec:level6}Conclusion}
In conclusion, we mitigate a major issue in active microrheology - that of extracting rheological parameters of a fluid over a large bandwidth with high signal-to-noise in a single shot. Thus, we propose a new technique, which we name as the \textit{'Multiple Sinusoids Superposition Method'}  -- where we use a superposition of multiple sine waves to excite a colloidal probe particle embedded in a PAM:water mixture at different concentrations to extract the complex rheological parameters. For low frequencies, we use a square wave - which is by definition a combination of odd harmonics of sines - while for higher frequencies - we apply a superposition of sine waves with the modulation amplitudes increasing with increasing excitation frequency in a proportional manner. This ensures that the signal-to-noise extracted at high frequencies does not diminish in spite of the low displacement of the probe particle at those frequencies. We use only the phase component of the probe response, which is advantageous since it does not require calibration of the displacement sensitivity of our detection system, and is not susceptible to spurious electronic noise which affects amplitude measurements considerably. We extract the phase from a discrete fast Fourier transform that we perform on the measured time series of the probe displacement, and obtain the complex rheological parameters by repeating the measurement at two different trap stiffnesses. Our extracted rheological parameters match reasonably well with values available in literature, which acts as a good consistency check. The method is fast, accurate, and is easily extendible to even higher frequencies by employing optical traps of higher stiffnesses and larger sampling rate for signal measurement. We are currently in the process of extending this method to microrheology experiments inside biological cells\cite{rigato2017high}, where it can possibly extract information at frequency values hitherto inaccessible by active microrheology. 
\medskip
\section*{SUPPLEMENTARY MATERIAL}
See Supplement file for supporting content.

%\section*{Data Availability Statement}
%The experimental data that support the findings of this study are available from the corresponding author upon reasonable request.

\begin{acknowledgments}
	This study was supported by IISER Kolkata, teaching and research institute supported by the Ministry of Human Resource Development, Govt. of India. A.K and R.D acknowledge the Department of Science and Technology, Govt. of India for the INSPIRE Fellowship and INSPIRE Scholarship respectively.
\end{acknowledgments}

%% Produces the bibliography via BibTeX.

	\begin{center}
		
		\textbf{\large Supplementary Information}
		\bigskip
		\bigskip
		
		{\bf Single shot wideband active mircorheology  using multiple-sinusoid modulated Optical Tweezers}\\
		\bigskip
		{\bf Avijit Kundu\textsuperscript{1}, Raunak Dey\textsuperscript{1}, Shuvojit Paul\textsuperscript{1,2}, Ayan Banerjee\textsuperscript{1}
		}
	
	\end{center}
	
	\setcounter{figure}{0}
	\section*{A. Amplitude response}
\begin{figure}[h!]
		\centering
		\includegraphics[height=6cm]{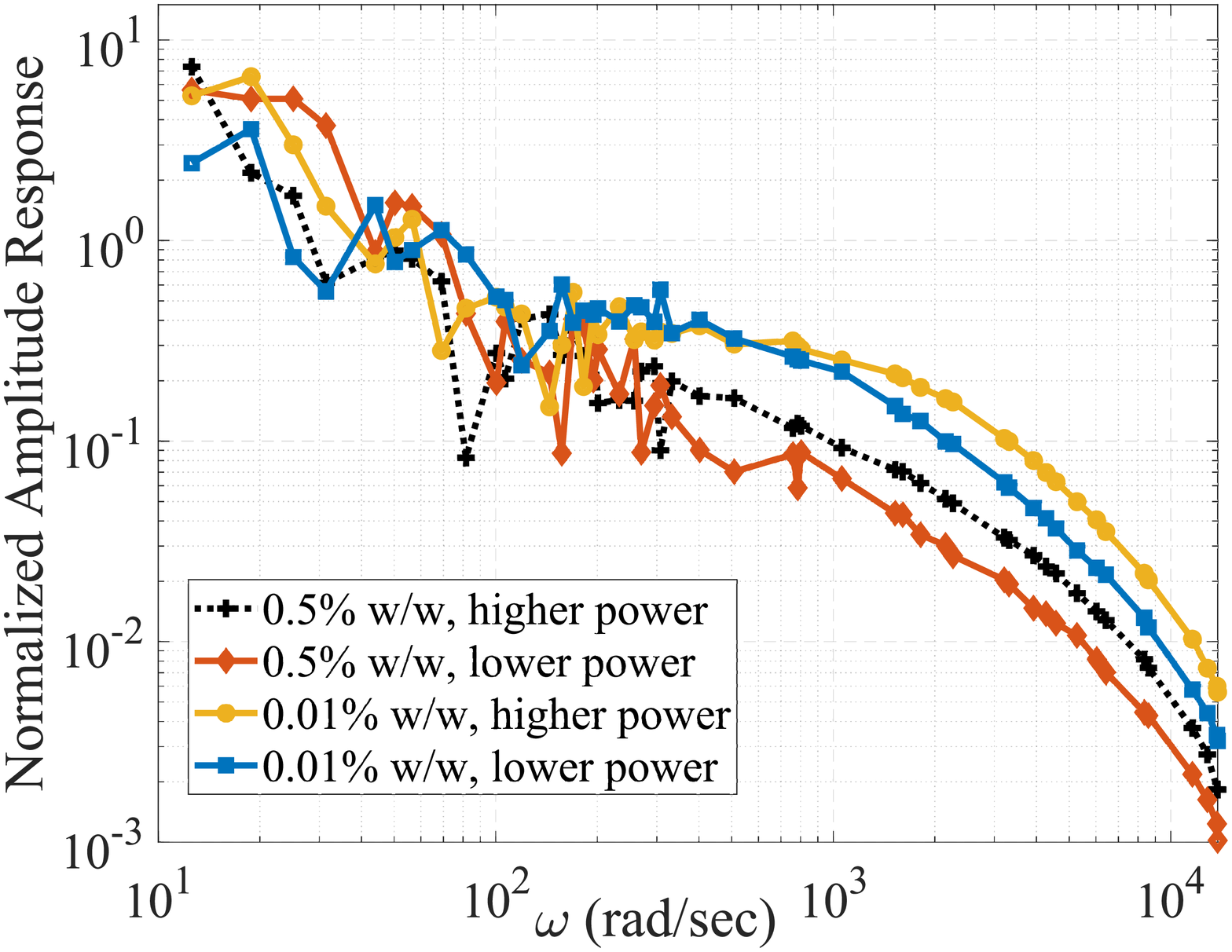}
		\caption{The normalized amplitude responses are obtained from the time series of particle displacement and plotted against the angular frequency of the active probe particle. Here the normalized amplitude means the ratio of the output signal to the input excitation for each frequency. We observe better amplitude response in samples with higher concentrations. We have normalised the curve, putting the maxima at 1, for clarity.}
		\label{fig:amp}
	\end{figure}

	Although we completely use the phase response from the trapped probe to compute the rheological parameters of the VE fluid, it principle this can also be done from the amplitude response -- which has its own drawbacks as discussed in the paper. But it is intriguing to note that the amplitude response of the particle is inversely proportional to the trapping power -- which is further corroborated by Eq.~10 of the main manuscript. The very increasing nature of the shear modulus with concentration is linked to the fact that the amplitude response is increases with increasing concentration of the samples. We demonstrate this in Fig.~\ref{fig:amp}, for two different concentrations and two different laser powers.
	
	\section*{B. Weissenberg Number}
	The Weissenberg number ($Wi$) which is defined as  $Wi=\frac{\dot{x}_o\lambda}{2a}$, where $\lambda$ is the relaxation time constant of the fluid and $a$ is the radius of the trapped colloidal probe. The method of obtaining the time constant is described in a previous work \cite{paul2019active}. The Weissenberg number is very commonly used in the literature to denote the ratio of the elastic to the viscous forces. 

		\section*{C. Storage modulus with modulation amplitude}
		The response of the viscoelastic system lies in the linear regime when $Wi<<1$ \cite{chapman2014nonlinear}. The Weissenberg number depends on the velocity of the trapped particle inside the solution. So we have modulated the particle with different amplitudes 70 nm and 130 nm respectively, and extracted the storage modulus of 1\% w/w PAM solution. At the higher modulation amplitude the Weissenberg number is close to 0.1,  where as for the lower amplitude it is close to 0.05. So performing the experiment at the higher amplitude modulation may introduce non-linearity in the response of the viscoelastic fluid at higher concentrations. We have plotted in Fig.~\ref{fig:G1_with_amp} both the storage moduli for modulating the particle with different amplitudes, and it shows that with higher amplitude-modulation the storage modulus does not saturate at high frequency as is the case for lower amplitude, since we probably approach the non-linear viscoelastic regime for the former. For this, we have used 70 nm modulation for the highest concentration. In future we will expand our study on the non-linear effects of viscoelastic fluids at higher concentrations.
	\begin{figure}[h!]
		\centering
		\includegraphics[height=6cm]{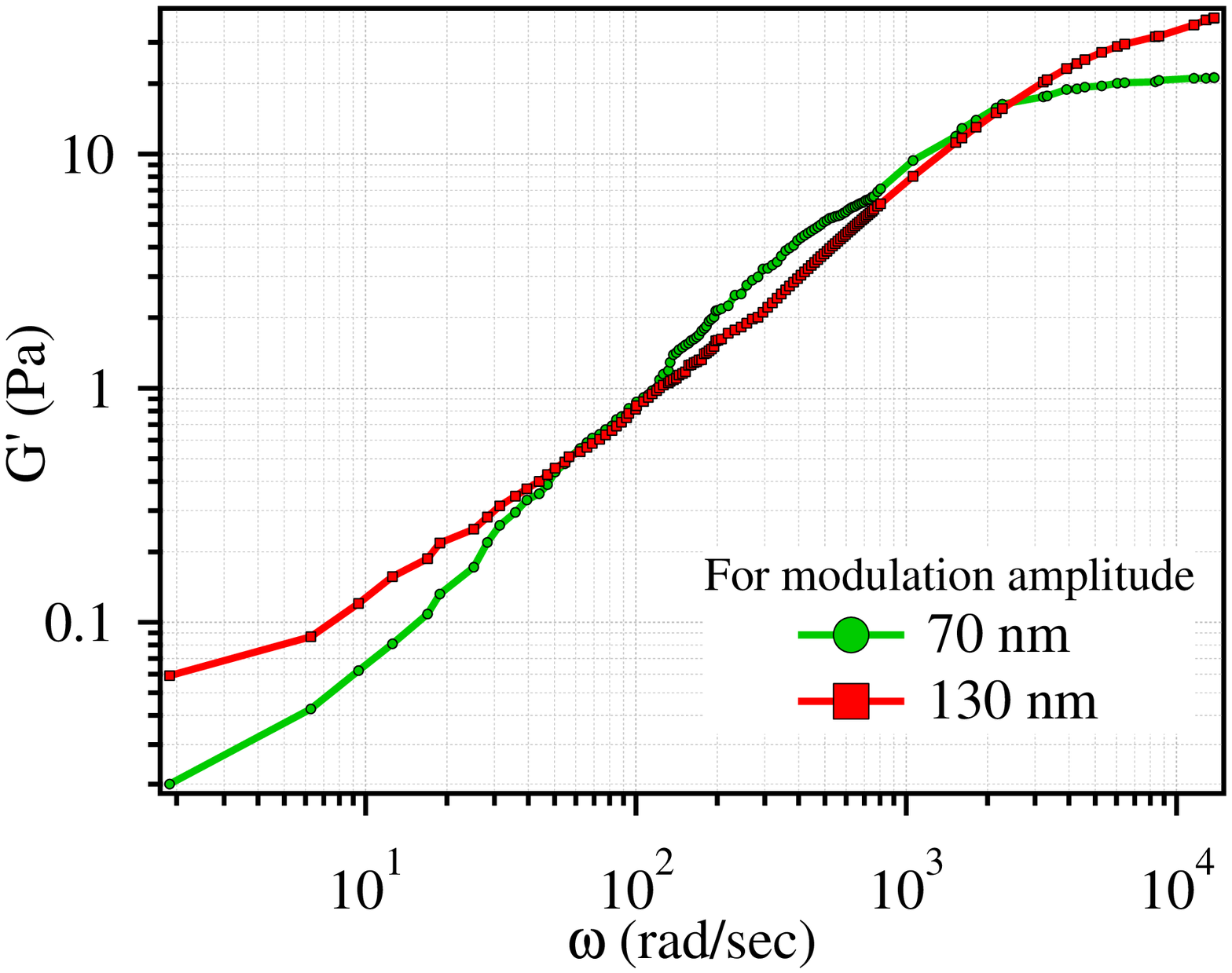}
		\caption{Storage moduli of 1\% w/w PAM solution extracted from modulating the particle by different amplitudes.}
		\label{fig:G1_with_amp}
	\end{figure}
	
	\section*{D. Inertia effect in viscoelastic fluid}
		We have neglected the inertia effect here as our measurement frequency range obeys the following relation according to Ref. 33 in the main manuscript.
		\begin{equation}
		\omega<<\sqrt{\pi^2G(\omega)/4a^2\rho_f} \label{eq:1}
		\end{equation}
		where $G(\omega)$ is the frequency dependent shear modulus, $a$ is the particle radius, and $\rho_f$ is the density of the particle. In our case the frequency is 100 krad/s and our measured frequency (14 krad/s is less than that. However at very high frequency there still may be some influence of the residual inertia of the fluid-probe system. Such effects can modify the saturation behaviour of the shear-moduli frequencies over 10 kHz frequency. Such effects are explicitly calculated in Ref.~\cite{grimm_se}.
		
		\section*{E. Viscoelastic moduli with error bar}
		\begin{figure}[h!]
		\centering
		\includegraphics[height=6cm]{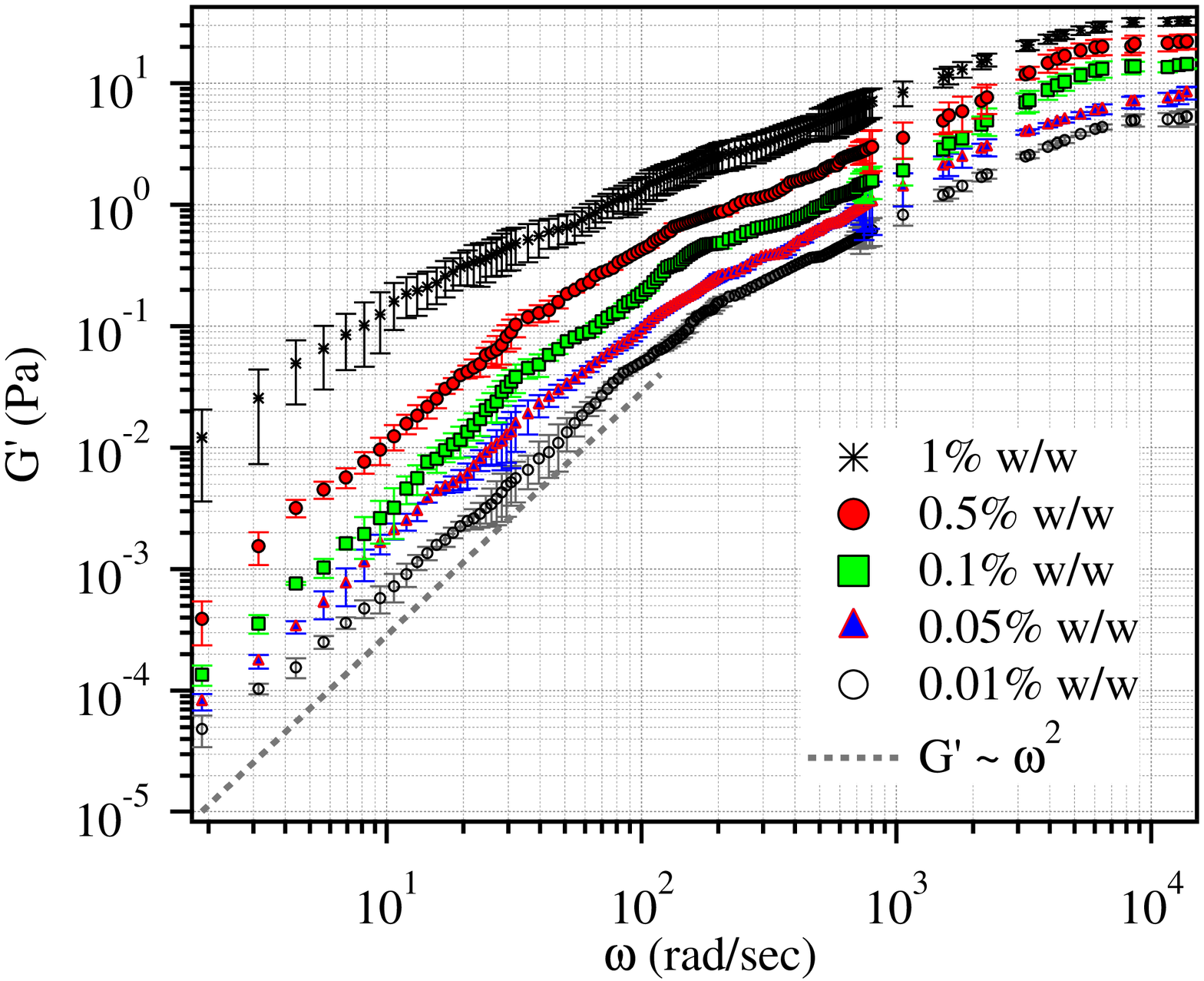}
		\caption{Storage moduli PAM solution in water for five different concentrations of 0.01\%, 0.05\%, 0.1\%, 0.5\%, 1\% w/w. Here storage moduli vary with $\omega^2$.}
		\label{fig:G1_with_amp}
	\end{figure}
	\begin{figure}[h!]
		\centering
		\includegraphics[height=6cm]{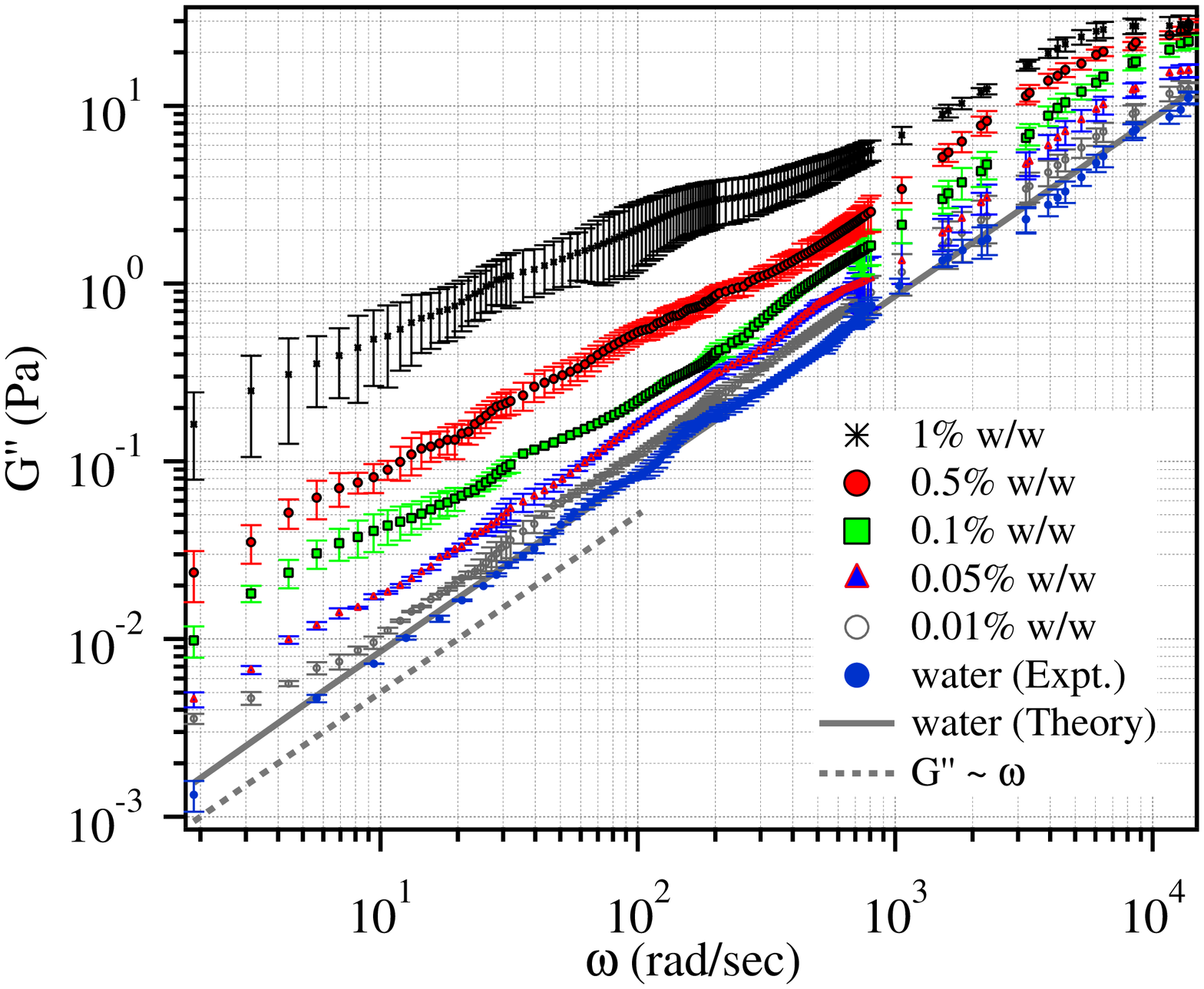}
		\caption{Loss moduli PAM solution in water for five different concentrations of 0.01\%, 0.05\%, 0.1\%, 0.5\%, 1\% w/w. Here the loss moduli vary with $\omega$.}
		\label{fig:G1_with_amp}
	\end{figure}

	%\nocite{*}
	%

\end{document}